\documentclass{iopjournal}

\usepackage{graphicx}
\graphicspath{{Figures/}}

\usepackage[normalem]{ulem}
\usepackage{wrapfig}
\usepackage{float}
\usepackage[utf8x]{inputenc}
\usepackage{amsmath}
\usepackage{amssymb}
\usepackage{amsthm}
\usepackage{mathtools}
\usepackage{color}
\usepackage{multirow}
\usepackage{tabularx}
\usepackage{orcidlink}

\usepackage{soul}
\usepackage{siunitx}
\usepackage{ulem}
\usepackage{booktabs}
\usepackage{ragged2e}

\usepackage{changes}

\DeclareSIUnit\sample{Sa}
\DeclareSIUnit{\bar}{bar}
\DeclareSIUnit{\pc}{pc}
\DeclareSIUnit{\kpc}{kpc}
\DeclareSIUnit{\dBm}{dBm}
\DeclareSIUnit{\Mpc}{Mpc}
\DeclareSIUnit{\weber}{Wb}
\DeclareSIUnit{\month}{month}
\DeclareSIUnit{\year}{year}
\DeclareSIUnit{\sqrthz}{\ensuremath{\sqrt{\text{Hz}}}}

\begin{document}
\justifying

\articletype{Paper}

\title{MAGGIE: A Magnetic Gravitational Wave Induction Experiment}

\author{Jasper Jödicke$^{1,*}$\orcidlink{0009-0003-5523-4111}, Marios Maroudas$^1$\orcidlink{0000-0003-1294-1433}, Toma-Stefan Cezar$^1$\orcidlink{0009-0009-5121-1821} and Dieter Horns$^1$\orcidlink{0000-0003-1945-0119}}

\affil{$^1$Institut für Experimentalphysik, Universität Hamburg, Luruper Chaussee 149, D-22761 Hamburg, Germany}
\newline
\affil{$^*$Author to whom any correspondence should be addressed.}

\email{jasper.joedicke@uni-hamburg.de}


\begin{abstract}

\leftskip=0pt \rightskip=0pt
Gravitational waves (GWs) can induce effective electromagnetic currents when interacting with external electric and magnetic fields, as described by linearized gravity modifications to Maxwell’s equations. This coupling enables a novel detection approach for high-frequency gravitational waves (HFGWs) using axion haloscope experiments. Here we present the MAGnetic Gravitational wave Induction Experiment (MAGGIE), the first lumped-element HFGW detector proposed in Europe, designed to probe HFGWs in the kHz-MHz regime by leveraging a \SI{14}{\tesla} solenoidal magnet at the University of Hamburg. The GW-induced magnetic flux is captured by a custom-designed pickup loop optimized for the expected symmetry of the effective current. A figure-8-shaped geometry, oriented to break the azimuthal symmetry, is implemented, together with a blind-loop configuration, for real-time noise rejection and calibration. The readout scheme is tailored for continuous signals and time-domain transient searches, using waveform templates for primordial black hole (PBH) mergers. The expected experimental reach in terms of strain spectral noise density at the \SI{40}{\MHz} high-frequency end is projected to reach $\sim 4 \times 10^{-16}/\sqrt{\mathrm{Hz}}$ for transient searches, and in terms of strain, projected to reach $\sim 10^{-19}$ for 1 year of continuous search. This allows MAGGIE to constrain currently unexplored regions of the HFGW parameter space.

\end{abstract}

\section{Introduction}
\label{sec:introduction}

The direct observation of gravitational waves (GWs) has emerged as one of the most transformative achievements in modern physics, opening a new window onto the Universe. Since the first detection by the LIGO interferometer in 2015 \cite{LIGO-main}, GW astronomy has enabled the study of compact binary mergers and tests of general relativity in the strong-field regime with unprecedented precision.

Ground-based interferometers such as LIGO \cite{LIGO-main}, Virgo \cite{VIRGO}, and KAGRA \cite{KAGRA} are primarily sensitive to GWs in the low-frequency range from \SI{1}{\Hz} to \SI{1}{\kHz} while in principle higher frequencies are accessible \cite{2025NatSR..1525733S}. However, a large portion of the GW spectrum remains unexplored, particularly in the high-frequency regime spanning from \si{\MHz} to \si{\GHz}. HFGWs are predicted in several beyond-Standard Model (BSM) scenarios, including early-universe phase transitions \cite{EarlyUniverse}, the dynamics of primordial black holes (PBHs) \cite{maggiore_2007}, neutron star oscillations \cite{Mateos_2022}, and superradiant instabilities involving ultralight bosons \cite{Aggarwal_2021}. The high-frequency GW background may also carry relic imprints from inflation or the Big Bang itself \cite{Beckwith_2009}.

Despite their theoretical appeal, detecting HFGWs is extremely challenging \cite{Aggarwal_2025}. Their feeble interaction with matter results in very weak signals, easily masked by thermal, quantum, and technical noise. Nonetheless, emerging experimental approaches achieve greater sensitivity at higher frequencies, especially those that use strong static magnetic fields. In particular, the Gertsenshtein effect \cite{JetP14.84.1962} predicts that GWs can induce electromagnetic signals when passing through a magnetic field, enabling their detection. Several existing and planned experiments exploiting strong-field environments can therefore be repurposed or extended to search for HFGWs \cite{GWDomcke}.

Here we present the MAGnetic Gravitational wave Induction Experiment (MAGGIE), a new experiment designed to search for HFGWs using a \SI{14}{\tesla} solenoidal magnet initially procured for the operation of axion haloscopes like the WISPLC experiment \cite{Zhang:2021bpa} and currently operated at the University of Hamburg. MAGGIE represents the first broadband, lumped-element HFGW detector proposed and under construction in Europe. The experiment employs a suitable pickup loop and readout chain to measure possible flux signals arising from the interaction of GWs with the magnetic field.

In Sect.~\ref{sec:theory}, we derive the theoretical magnetic flux induced in the pickup loop due to the coupling of a GW to the magnetic field. Sect.~\ref{sec:experiment} discusses the symmetry properties and optimal geometry of the pickup loop configuration. The data acquisition and readout system are described in Sect.~\ref{sec:daq}, followed by a preliminary estimate of the experimental sensitivity in Sect.~\ref{sec:sensitivity}. We contextualize MAGGIE within the broader experimental landscape in Sect.~\ref{sec:discussion}, and finally, discuss strategies for sensitivity improvement and prospects for HFGW detection in Sect.~\ref{sec:outlook_conclusion}.

\section{Theoretical Background}
\label{sec:theory}

\begin{figure}[t!]
    \centering
    \includegraphics[width=0.6\linewidth]{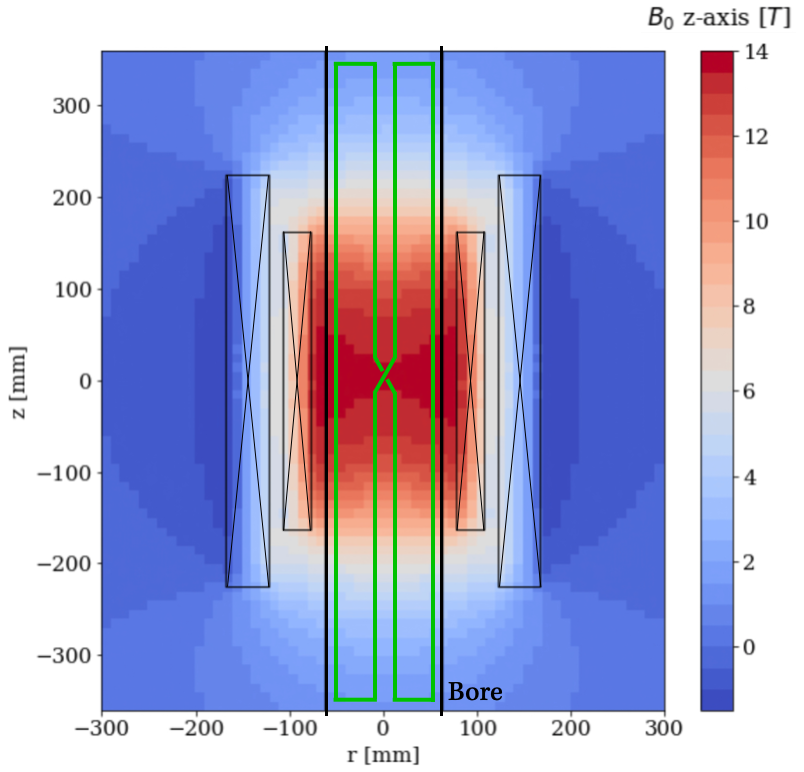}
    \caption{Simulated magnetic field distribution of the \SI{14}{\tesla} warm-bore solenoidal superconducting magnet used for MAGGIE, calculated using COMSOL. The field is azimuthally symmetric and varies with radial and axial position. The bore volume is indicated in black, and the sensitive measuring loop in green.}
    \label{fig:Bext}
\end{figure}

In the presence of an external magnetic field, shown in Fig. \ref{fig:Bext}, a passing GW can induce an effective displacement current via its coupling to electromagnetism in curved spacetime \cite{JetP14.84.1962,PhysRevD.37.1237, Berlin_2021}. We begin with flat Minkowski spacetime $\eta_{\mu \nu}$ and consider a linear perturbation $g_{\mu \nu} = \eta_{\mu \nu} + h_{\mu \nu}$, where $h_{\mu \nu}$ encodes the GW with its two polarizations $h^+$ and $h^\times$. In this background, Maxwell’s equations are modified as:
\begin{equation}
    \partial_\nu F^{\mu \nu} = j^\mu +  j^\mu_{\text{eff}},
\end{equation}
where 
$j^\mu$ is the electromagnetic current without the presence of GWs, $F^{\mu \nu}$ is the electromagnetic field tensor, and $j^\mu_{\text{eff}}$ the induced effective displacement current which is given by \cite{GWDomcke, GWDomcke2}:
\begin{equation}
    j^\mu_{\text{eff}} = \partial_\nu \Big (-\frac{h}{2}F^{\mu \nu}+F^{\nu}_{\hspace{5pt} \alpha}h^{\alpha \mu} + F^{\mu}_{\hspace{5pt} \beta}h^{\beta \nu}\Big).
\end{equation}

The effective current acts as a source for a secondary magnetic field, which can be computed using the Biot–Savart law:
\begin{equation}
    \boldsymbol{B}_{\mathrm{ind}} (\boldsymbol{r},t) = \frac{1}{4\pi}\int\limits_{V_B} \mathrm{d^3}r' \, \frac{\textbf{j}_{\text{eff}}(\boldsymbol{r'},t) \times (\boldsymbol{r}-\boldsymbol{r'})}{|(\boldsymbol{r}-\boldsymbol{r'})|^3},
\end{equation}
where the integration is performed over the volume of the magnet bore $V_B$ in cylindrical coordinates $\boldsymbol{B}_{\mathrm{ind}} (\boldsymbol{r}) = B_r \, \hat e_r + B_\phi \, \hat e_\phi + B_z \, \hat e_z$. 

By positioning a suitably oriented pickup loop within the bore, the induced magnetic field passes through the loop, generating a magnetic flux $\Phi_{\text{ind}}$ whose change can be measured as an electrical current. The magnetic flux is dependent on the pickup-loop area $A_l$ and is given by
\begin{equation}
     \Phi_{\text{ind}}(t) = \int\limits_{A_l} \mathrm{d} \boldsymbol{F} \cdot \boldsymbol{B}_{\text{ind}}(\boldsymbol{r}, t).
\end{equation}

Alternatively, for a given pickup-loop geometry, the flux can be computed from the vector potential $\boldsymbol{A}(\boldsymbol{r})$ using a line integral:
\begin{equation}
    \Phi_{\text{ind}}(t) = \oint\limits_\gamma \mathrm{d}\boldsymbol{s} \cdot \boldsymbol{A}(\boldsymbol{s}, t) =  \oint\limits_\gamma \,  \mathrm{d}\boldsymbol{s} \cdot \int\limits_{V_B} \mathrm{d}^3r' \, \frac{\textbf{j}_{\text{eff}}(\boldsymbol{r'},t)}{|(\boldsymbol{s}-\boldsymbol{r'})|}.
    \label{eq:PHImain}
\end{equation}

\begin{figure}[h!]
    \centering
    \includegraphics[width=0.5\linewidth]{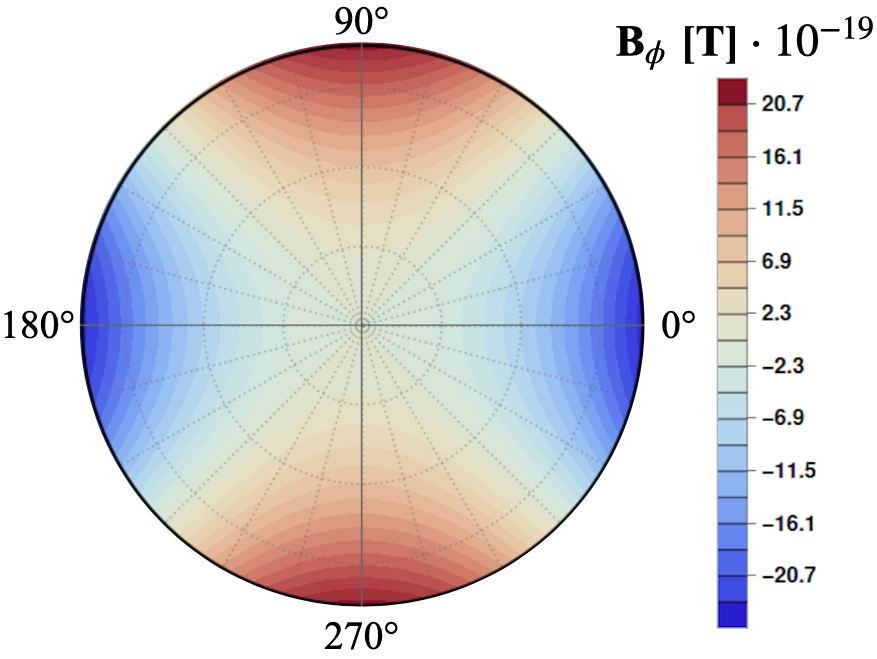}
    \caption{Distribution of the magnetic field density induced by a GW. Shown is the azimuthal $\phi$-component, $B_\phi$, for a purely $\times$-polarized GW at $\SI{40}{\mega\hertz}$ with strain $h_\times = 10^{-17}$ and incoming angles $(\theta_h = \pi/2, \phi_h= \pi)$ chosen to maximize the coupling. The radial coordinate denotes the distance from the central longitudinal axis, spanning the full \SI{125}{\mm} bore diameter.}
    \label{fig:currentdis}
\end{figure}

As will be discussed in Sect.~\ref{sec:experiment}, the effective current and resulting induced magnetic flux exhibit a $\cos(2\phi)$-symmetric distribution in the leading order of the GW frequency $\omega$, as shown in Fig.~\ref{fig:currentdis}.

This lowest-order contribution dominates in the magneto-quasistatic limit where $\omega L \ll 1$, with $L$ characterizing the detector’s spatial extent and using $c=\epsilon_0 = \mu_0=1$. Exploiting this symmetry, the pickup loop is designed with a figure-8 geometry, consisting of two adjacent rectangular loops that carry opposing induced currents. For such a configuration, the induced flux from Eq.~\ref{eq:PHImain} can be expressed as \cite{GWDomcke2}:
\begin{equation}
    \Phi_8^{(2)} = \frac{e^{-i\omega t}}{36\sqrt{2}} \omega^2 B_0 l r (30R^2-13r^2)\sin{\theta_h}(h_\times \cos{\phi_h}+h_+ \cos{\theta_h}\cos{\phi_h}),
    \label{eq:8PHI}
\end{equation}
where $\phi_h$ and $\theta_h$ denote the angles of the incoming GW, $\phi_l$ is the placement angle of the pickup-loop and set to $0$, $B_0$ is the external field magnitude, $R$ the radius of the bore, $l$ the length and $r=r_2-r_1$ the radial extension of the pickup-loop. As will be demonstrated in Sect.~\ref{sec:experiment}, the figure-8 loop geometry maximizes sensitivity to the GW signal under these symmetry conditions.

Detecting these induced signals in the \si{\kHz} to \si{\GHz} range opens up new possibilities for probing BSM physics. Potential astrophysical and cosmological sources include early-universe phenomena that contribute to a broadband stochastic GW background \cite{Christensen_2019}, transient mergers of PBHs \cite{maggiore_2007}, monochromatic or broadband signals from superradiant instabilities around rotating black holes \cite{Aggarwal_2025}, as well as neutron star mergers and other compact binary coalescences \cite{Mateos_2022}. Table~\ref{tab:GWsources} summarizes the expected frequency ranges and characteristic strain amplitudes for several of these key source classes, establishing the target parameter space for the MAGGIE detector.

\begin{table}[h]
    \centering
    \begin{tabular}{lcc}
        \toprule
        \textbf{Source} & \textbf{Frequency Range} & \textbf{Strain} $h(f)$ \\
        \midrule
        \textit{Transient Signals:}\vspace{0.2cm} & & \\
	PBH Mergers \cite{maggiore_2007} & $10^4$ -- $10^9$ Hz &\hspace*{-3.5pt}$10^{-22}$ -- $10^{-27}$ \\
        NS Mergers \cite{Mateos_2022}  & $10^3$ -- $10^7$ Hz & $10^{-22}$ -- $10^{-26}$ \vspace{0.3cm} \\
        \textit{Continuous Signals:} \vspace{0.2cm} & & \\
        Superradiance \cite{Aggarwal_2021}  & $10^4$ -- $10^9$ Hz & $10^{-18}$ -- $10^{-25}$  \\
        Early Universe \cite{Christensen_2019} & $10^3$ -- $10^9$ Hz & $10^{-25}$ -- $10^{-30}$ \\
        \bottomrule
    \end{tabular}
    \caption{Overview of potential HFGW sources and their expected strain amplitudes. All strains $h(f)$ are estimated for source distances of \SI{1}{\kpc} for compact object mergers (PBHs, neutron stars) and superradiance, and for horizon-scale distances in the case of early-universe signals.}
    \label{tab:GWsources}
\end{table}

\section{Experimental Design}
\label{sec:experiment}

\begin{figure}[!htb]
    \centering
    \includegraphics[width=\linewidth]{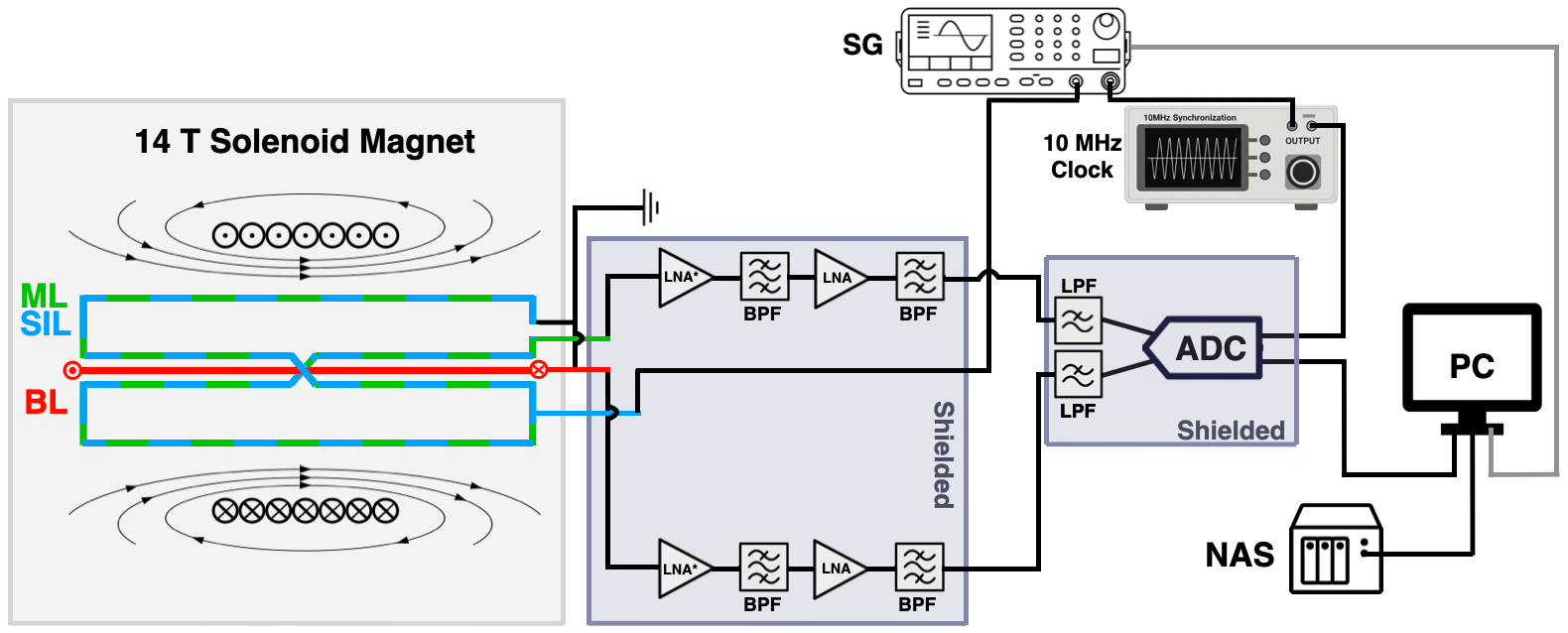}
    \caption{Schematic overview of MAGGIE. The \SI{14}{\tesla} solenoidal magnet houses three pickup loops: the measurement loop (ML, green), the signal injection loop (SIL, blue) positioned adjacent to the ML with the same geometry, and the blind loop (BL, red) rotated by $90^\circ$. Signals from the ML and BL are read out by high-input-impedance amplifiers (LNAs$^*$), passed through band-pass filters (BPFs), further amplified by standard low-noise amplifiers (LNAs), filtered by an additional narrow-band BPF, and conditioned via an anti-aliasing low-pass filter (LPF) before being digitized by a two-channel analog-to-digital converter (ADC) located in a separate shielded enclosure. A signal generator (SG) injects calibration signals via the SIL.}
    \label{fig:schematic}
\end{figure}

A schematic overview of the MAGGIE setup is shown in Fig.~\ref{fig:schematic}. The experiment employs a \SI{14}{\tesla} solenoidal magnet to search for HFGWs. The superconducting coils are maintained at \SI{4}{\kelvin}, while the central bore remains at room temperature, providing convenient access for installing and calibrating the detector system. The bore has a diameter of \SI{125}{\mm} and a length of \SI{755}{\mm}, and the magnetic field is azimuthally symmetric, depending on the radial and axial coordinates, $B_{\text{ext}} = B_{\text{ext}}(r, z)$, as shown in Fig.~\ref{fig:Bext}.

Inside the warm bore, three pickup loops are mounted within aluminum shielding to suppress external electromagnetic interference (EMI). These comprise the sensitive measurement loop, the symmetric blind loop, and the signal injection loop (see Sect.~\ref{sec:loops}). The measurement and blind loop signals are routed through feedthroughs into a dedicated shielded electronics enclosure, where they are read out by high-input-impedance amplifiers for direct voltage detection. This approach minimizes current-noise contributions and maximizes the detector's sensitivity to measured induced voltage. It is noted that the upper frequency limit of the experiment is constrained by the parasitic capacitance of the pickup loop and cabling. At higher frequencies, this stray capacitance acts as a low-pass filter, shunting the signal to ground and heavily attenuating the induced voltage. Based on experience from the WISPLC \cite{Zhang_2022_wisplc} prototype setup, the system is expected to maintain a linear response up to about \SI{40}{\MHz}, beyond which capacitive effects begin to dominate.

Following the initial amplification, the signal is passed through band-pass filters (BPFs) to eliminate low-frequency environmental noise and to reject high-frequency out-of-band interference. A second amplification stage using low-noise amplifiers (LNAs) is then followed by an additional narrow-band BPF to further suppress out-of-band artifacts. Immediately before digitization, a final low-pass filter with a cutoff of approximately \SI{40}{\MHz} provides anti-aliasing protection, ensuring that no spectral folding contaminates the analysis band. The conditioned outputs are then fed into a two-channel analog-to-digital converter (ADC), located in a separate shielded enclosure to avoid interference from other electronics. This setup enables simultaneous digitization of the measurement and blind loop signals, allowing real-time background rejection and noise diagnostics. Such extensive shielding and filtering are crucial in a broadband HFGW search, where sensitivity is ultimately limited by even weak external noise.

Additionally, a signal generator (SG) is used to inject well-defined calibration signals into the system via the signal injection loop (SIL), enabling end-to-end verification of the readout chain and the matched-filter analysis (see Sect.~\ref{sec:daq}). Both the SG and the ADC are phase-locked to a \SI{10}{\MHz} reference clock to ensure precise synchronization during calibration and data acquisition.

Finally, digitized data are continuously streamed to a network-attached storage system. As will be discussed in Sect.~\ref{sec:daq}, to manage the large data volume, only selected outputs are permanently recorded for offline analysis: candidate triggers and noise diagnostics for transient searches, and time-averaged power spectra for continuous signal searches.

\subsection{Induced Magnetic Flux}
\label{sec:induced_flux}

The response of the pickup loop, which determines the expected signal, is governed by the magnetic flux induced by the interaction of the external field with a passing HFGW given by Eq.~\ref{eq:PHImain}. Since the external magnetic field is oriented in the $z$-direction, only the $\rho$ and $\phi$ components of the secondary magnetic flux density are accessible. Among these, the $\phi$ azimuthal component provides the simplest measurable signal. The dominant contribution arises from the effective current density, and the spatial distribution of the magnetic field density $B_\phi$ is shown in Fig.~\ref{fig:currentdis}.

This $\cos 2\phi$ - symmetric shape in the current distribution translates into a symmetric flux pattern through a loop and must be broken by the pickup geometry to achieve nonzero net flux. Two loop configurations can achieve this: either an array of loops arranged vertically along one side of the bore, parallel to the external field, or an array of loops spanning the full diameter of the bore, with a reversal of current direction at the center. The latter results in a figure-8 configuration. The induced flux for such a geometry simplifies to the following expression:
\begin{equation}
    \Phi_{\text{ind}} = \int\limits_{r_1}^{r_2} \mathrm{d}r \ 2 A_r (r, \phi_l, l/2) +
    \int\limits_{-l/2}^{l/2} \mathrm{d}z \, \left(A_z(r_1, \phi_l, z)
    - A_z (r_2, \phi_l, z)\right),
    \label{eq:param}
\end{equation}
where $r_2-r_1$ is the radial size, $l$ the total length, and $\phi_l=0$ the angular placement of the loops. However, for the specific magnetic field profile shown in Fig.~\ref{fig:Bext}, the integral must be evaluated numerically.

\subsection{Signal Injection and Blind Loops}
\label{sec:loops}

A key feature of MAGGIE is its multi-loop architecture, which enables robust detection, noise characterization, and in situ calibration. This is accomplished using three distinct pickup loops: a sensitive measurement loop, a symmetric blind loop, and a signal injection loop. While the measurement loop is optimized to detect possible HFGW-induced flux signals, the blind and injection loops serve critical supporting roles.

The blind loop is specifically constructed to cancel out any magnetic flux that would result from a GW signal. It achieves this by mirroring the azimuthal distribution of the effective current (see Fig.~\ref{fig:currentdis}) and preserving the system's cylindrical symmetry (resembling an O-shape). As a result, any GW-induced flux contributions from opposite sides of the loop cancel out, yielding zero net signal. This makes the blind loop a powerful diagnostic tool for background rejection. When the external magnetic field is off, the blind loop serves as a reference channel for identifying common-mode noise, such as vibrational resonances or EMI. When the field is on, it serves as a veto: any coincident signals in both loops can be excluded as non-physical and can be confidently attributed to noise. The blind loop is positioned orthogonal to the measurement loop (rotated by $90^\circ$) to minimize inductive coupling and ensure an independent response. In future frequency-domain searches for persistent or narrowband signals, the blind loop can also serve to characterize the background noise floor. Peaks appearing in both loops can be explicitly flagged as EMI. This coincidence-matching allows for a clear distinction between environmental artifacts and true continuous GW candidates, suppressing residual systematics and improving overall sensitivity (see Sect.~\ref{sec:daq}).

The signal-injection loop is placed parallel to the measurement loop, offset laterally by $\sim\SI{0.5}{\cm}$, and wound in the same figure-8-shaped geometry. It enables controlled calibration of the readout chain by inductively injecting well-defined signal templates. This allows characterization of the system's timing, gain, and phase response, as well as end-to-end verification of signal recovery algorithms. This design ensures efficient coupling between the injection and measurement loops via mutual inductance.

\begin{figure}[!htb] 
    \centering
    \includegraphics[width=0.3\linewidth]{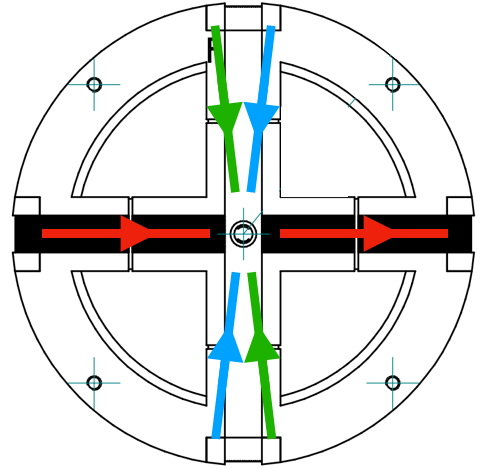}
    \includegraphics[width=0.3\linewidth,angle=90]{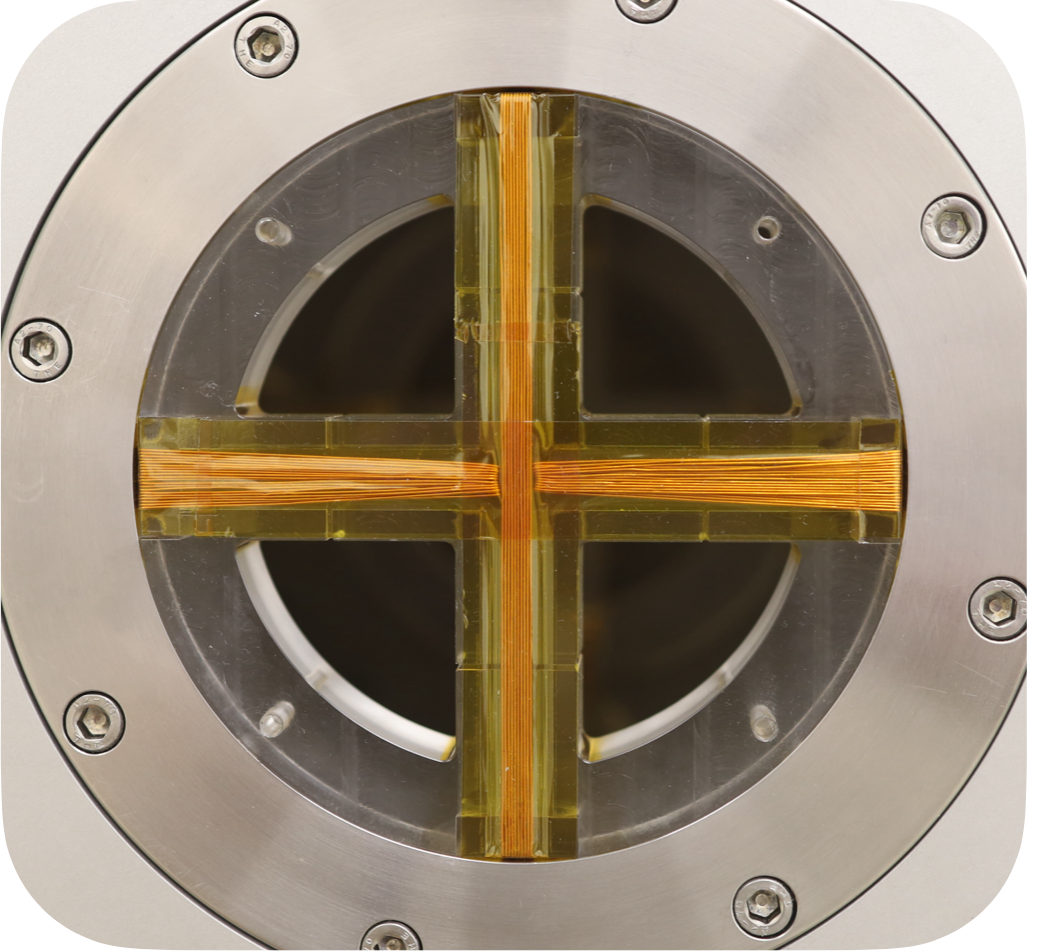}
    \includegraphics[width=0.65\linewidth]{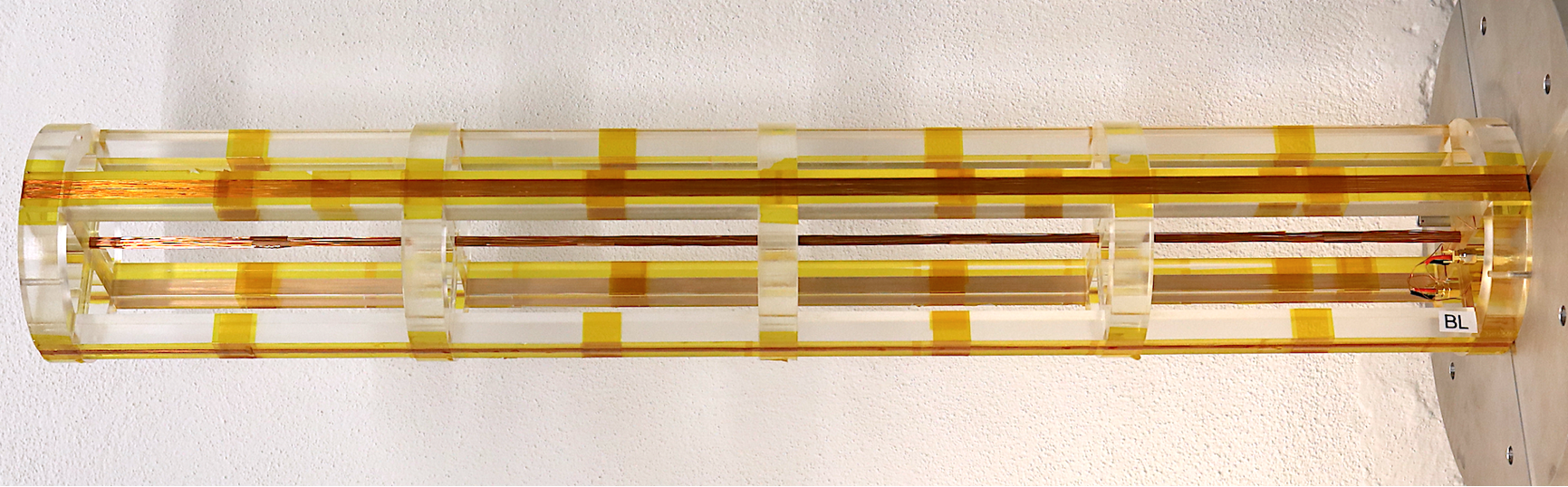}
    \caption{Overview of the pickup loop geometry, the frame, and installed windings. \textbf{Top Left:} The loop geometry of the final setup with an 8-shaped measurement loop (green), a blind loop (red), and a signal-injection loop (blue). Arrows indicate the direction of current (wire winding). \textbf{Top Right:} 
    First prototype of the installed loops with the optimal loop geometry to maximize the net signal from the GW effective current. The apparatus is inserted into the warm bore of the \SI{14}{\tesla} magnet at the University of Hamburg. \textbf{Bottom:} Initial prototype of the complete pickup loop frame.}
    \label{fig:looptopos}
\end{figure}

For the first constructed prototype, which serves as an initial baseline and remains under active optimization, each loop consists of ten windings of copper wire with a diameter of \SI{0.4}{\mm} and a spacing of \SI{0.05}{\cm} between adjacent turns. The mutual alignment and winding geometry of the loops are illustrated in Fig.~\ref{fig:looptopos}.

\subsection{Data Acquisition and Readout Scheme}
\label{sec:daq}

The MAGGIE data acquisition system is designed for continuous, high-throughput recording and real-time GW signal analysis. A custom software pipeline handles real-time streaming, filtering, and evaluation of data, tailored to the unique signal morphology expected in this experiment. The signal readout from the two pickup loops (sensitive and blind) is digitized using an ADC card following a recent implementation in \cite{Cezar_2026_modular}.

Given the large volume of data, storing the full raw data is impractical. Instead, the system employs a matched-filtering algorithm in real time to identify signal candidates. Only relevant information, such as candidate triggers and noise diagnostics, will be saved for post-processing. The matched-filtering pipeline is built using CUDA/C code to leverage NVIDIA GPUs for high-speed computation. It uses a template bank generated using the \texttt{pycbc} Python package \cite{pycbc}. Data will be transferred directly between the digitizer and GPU via PCIe x16 using Remote Direct Memory Access (RDMA), ensuring minimal latency and high bandwidth. This architecture enables efficient filtering over a large template bank spanning various signal parameters (e.g., frequency, duration, source mass).

By using a matched-filtering approach, the isolated GW signatures in the time-domain data can be enhanced to achieve a higher signal-to-noise ratio (SNR). Assuming a signal $s(t) = h(t) + n(t)$ consisting of a known GW signature $h(t)$ and a noise component $n(t)$ with known strain-equivalent spectral noise density (SND) $\sqrt{S_h(f)}$, the optimal filter is defined as \cite{moore-2014}:
\begin{equation}
    \tilde K (f) = \frac{\tilde h(f)}{S_h (f)},
\end{equation}
where $\tilde h(f)$ is the Fourier transform of $h(t)$. The optimal SNR $\rho_{\mathrm{optimal}}$, assuming a perfect match between the observed GW signal and the applied filter template, is then given by:
\begin{equation}
    \rho_{\mathrm{optimal}} = \sqrt{\langle h | h \rangle},
\end{equation}
where the inner product $\langle\cdot |\cdot \rangle$ is defined as:
\begin{equation}    
    \langle A | B \rangle = 4 \Re\left\{ \int\limits_{0}^{\infty} \mathrm{d}f \frac{\tilde A^* (f) \tilde B(f)}{S_h(f)} \right\}.
\end{equation}

However, due to limited computational resources and the unknown parameters of the incoming GW signal, it is only possible to probe the measured time-domain signal against a fixed template bank. This introduces a mismatch between the incoming signal $h_{\mathrm{meas}}$ and the applied filter template $h_{\mathrm{t}}$. Therefore, the SNR becomes:
\begin{equation}
\label{eq:SNRMF}
    \rho_{\mathrm{mismatch}} = \frac{\langle h_{\mathrm{t}} | h_{\mathrm{meas}} \rangle}{\sqrt{\langle h_{\mathrm{t}}| h_{\mathrm{t}}\rangle}}.
\end{equation}

Rewriting Eq.~\ref{eq:SNRMF} using the convolution theorem gives the following time-series SNR, where $t$ is the time offset between the applied filter template and the measured GW signal:
\begin{equation}
\begin{aligned}
\label{eq:SNRMF_final}
    \rho_{\mathrm{mismatch}}(t) = \frac{4}{\sqrt{\langle h_{\mathrm{t}}| h_{\mathrm{t}}\rangle}} \, \Re\left\{ \mathcal{F}^{-1}\left\{\frac{\tilde h_\mathrm{t}}{S_h}\right\} \star h_{\mathrm{meas}} \right\}(t).
\end{aligned}
\end{equation}
By computing the cross-correlation ($\star$) of the measured time-domain signal $h_{\mathrm{meas}}$ with precomputed templates $\mathcal{F}^{-1}\{\tilde h_\mathrm{t}/S_h\}$, we can improve our sensitivity and enable real-time candidate detection, which can be used for selective data storage. A more detailed discussion can be found in Appendix \ref{app:matchedfilter}.

Matched-filtering systems such as \texttt{pycbc} \cite{pycbc} have shown excellent performance in the LIGO/Virgo band \cite{nitz-2018}, but their behavior in the high-frequency regime remains to be fully validated. The MAGGIE pipeline is designed to evaluate and benchmark this performance under the unique constraints of \si{\MHz}-scale GW signals.

\section{Sensitivity Estimation}
\label{sec:sensitivity}

The sensitivity of MAGGIE is determined by several key factors, including the geometry and orientation of the pickup loop, the calibration and signal injection strategy, and the computational resources available for real-time data analysis.

The geometry and orientation of the pickup loop directly impact the induced magnetic flux from GWs. Both the number of windings and the spatial coverage of the loop array contribute to enhancing sensitivity. For the sensitivity projections presented here, the proposed baseline configuration assumes ten figure-8-shaped windings optimized for the \SI{14}{\tesla} solenoidal field. \footnote{In the following, SI units are used.}

Using the known magnetic field profile and the pickup loop’s geometry, the induced magnetic flux from a plane GW with strain amplitude $h_\times$ can be numerically calculated:
\begin{equation}
    \Phi_{\text{ind}} = 8.86 \times 10^{-3}~\si{\weber} \, \cos{(\omega t)} \, h_\times \sin \theta_h \cos \theta_h \cos \phi_h   \left( \frac{f}{\SI{40}{\mega\hertz}} \right)^2\left( \frac{\eta}{0.515} \right)\left( \frac{B_0}{\SI{14}{\tesla}} \right)\left( \frac{N}{10} \right) \left( \frac{A_l}{\SI{914}{\cm}^2} \right),
    \label{eq:phifinal}
\end{equation}
where $\omega = 2\pi f$ is the angular frequency, $\phi_h, \theta_h$ the incoming angles of the GW, $f$ the frequency, $B_0$ the external magnetic field, $\eta$ the form factor accounting for the effective magnetic field across the loop area, $N$ the number of loops, and $A_l$ the area of one loop. In the following, the angles are chosen to maximize the coupling, i.e., $\theta_h = \pi/2, \phi_h = \pi$. The prefactor encapsulates the integrated effect of loop geometry and field configuration. Differentiating with respect to time gives the induced voltage:
\begin{equation}
\label{eq:Uind}
    V_{\text{ind}} = -\partial_t \Phi_{\text{ind}} = 1.11\times 10^6~\si{\volt}\, \sin{(\omega t)} \, h_\times \, \left( \frac{f}{\SI{40}{\mega\hertz}} \right)^3\left( \frac{\eta}{0.515} \right)\left( \frac{B_0}{\SI{14}{\tesla}} \right)\left( \frac{N}{10} \right) \left( \frac{A_l}{\SI{914}{\cm}^2} \right),
\end{equation}
exhibiting the characteristic cubic frequency scaling and a $90^\circ$ phase shift between the incoming GW and the resulting voltage signal.

\subsection{Continuous Signal Sensitivity}
For continuous signals, such as superradiance, relics from the early universe, or the GW background, the sensitivity is set by the strain-equivalent SND and by the integration time. For a stationary signal at frequency $f$, the strain-equivalent SND follows directly from the measured voltage noise and the calibrated strain-to-voltage response of the pickup loop:
\begin{equation}
    \sqrt{S_h} = 4.37\times10^{-16}\,
    \frac{1}{\sqrt{\si{\hertz}}}
    \left( \frac{\SI{40}{\MHz}}{ f} \right)^3
    \left( \frac{0.515}{\eta} \right)
    \left( \frac{\SI{14}{\tesla}}{B_0} \right)
    \left( \frac{10}{N} \right) 
    \left( \frac{\SI{914}{\cm}^2}{A_l} \right) 
    \left( \frac{T_\mathrm{sys}}{\SI{350}{\kelvin}}\right)^{\frac{1}{2}}
    \left( \frac{3}{\mathrm{SNR}} \right),
    \label{eq:strain_sensitivityCS}
\end{equation}
where $T_{\mathrm{sys}}$ is the total system noise temperature\footnote{This system temperature corresponds to a voltage SND of $\sqrt{S_{V,\mathrm{thm}}}=\SI[per-mode=symbol]{4.91e-10}{\volt\per\sqrthz}$.} accounting for the detector's physical temperature and the LNA's noise temperature. Coherent averaging over $N_{\mathrm{avg}}$ spectra with a resolution bandwidth $\Delta f$ results in a minimum detectable continuous strain of:
\begin{equation}
    h_\mathrm{min}^{\text{cont}}(f) = \sqrt{S_h} \sqrt{\frac{\Delta f}{N_{\mathrm{avg}}}}.
\end{equation}

\begin{figure}[ht!]
    \centering        
    \includegraphics[width=0.7\linewidth]{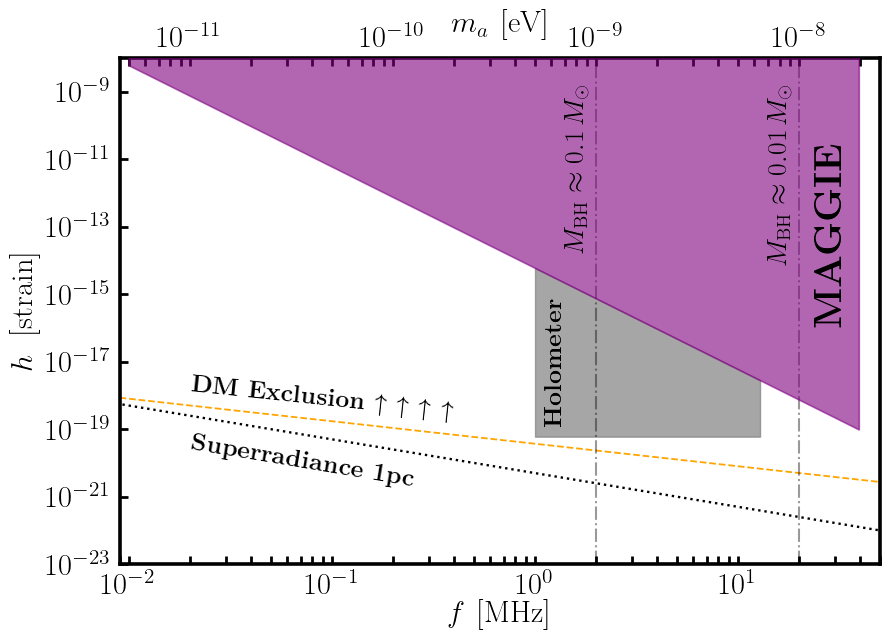}
    \caption{Projected strain sensitivity of MAGGIE across the \SI{10}{\kHz} to \SI{40}{\MHz} frequency band for a $\times$-polarized continuous GW signal, assuming an integration time of \SI{1}{\year} while accounting for a 30\% time-averaged geometric acceptance reduction due to Earth's rotation. At the high-frequency limit of \SI{40}{\MHz}, which corresponds to an axion mass of $m_a=\SI{2e-8}{\electronvolt}$ \cite{Arvanitaki1}, the projected minimum detectable strain reaches $\sim 10^{-19}$. The latest Holometer limit \cite{Holometer} along with the expected strain from black hole superradiance from \cite{Arvanitaki2} at a reference distance of \SI{1}{\pc} are shown for comparison. The orange line indicates the exclusion bound derived from local dark matter (DM) density constraints (see Appendix~\ref{app:ExclDM}). Experimental parameters used for this projection are summarized in Table~\ref{tab:ExpParas}.}
    \label{fig:sens-continuous}
\end{figure}

The projected strain sensitivity for continuous signals is shown in Fig.~\ref{fig:sens-continuous} for an integration time of \SI{1}{\year}, assuming a SNR of $3$. As expected, the sensitivity improves with increasing coherent integration time according to $h_{\mathrm{min}} \sim T^{-1/2}_{\mathrm{meas}}$. The intrinsic bandwidth of the signal is set by the GW emission timescale, which is determined by the model of gravitational coupling $\alpha$ and the superradiant process \cite{Brito_2020}.

Taking a second as an upper limit sets the bandwidth to $\Delta f=\SI{1}{\hertz}$. For longer-lived signals, however, this narrow intrinsic width is, in practice, washed out by Doppler modulation from Earth's orbital motion and rotation, which broadens the observed bandwidth well beyond \SI{1}{\hertz} as the integration time approaches a year. Furthermore, because the pickup loop possesses a highly directional antenna pattern, where the coupling efficiency depends strongly on the incoming angles $\theta_h$ and $\phi_h$ (see Eq.~\ref{eq:phifinal}), Earth's rotation introduces a time-dependent geometric signal reduction. As the detector's orientation relative to a fixed celestial source changes continuously over a sidereal day, the signal undergoes daily amplitude modulation. To account for this time-averaged geometric acceptance, a conservative 30\% reduction in the effective strain sensitivity is included for observation times exceeding one day. Future data analysis pipelines will include full directional demodulation to correct for Earth's rotational and orbital Doppler shifts, recovering a narrow bandwidth, reducing the integrated noise, and properly including the geometric antenna pattern.

As a representative benchmark, Fig.~\ref{fig:sens-continuous} includes the expected strain from superradiant emission at a characteristic distance of \SI{1}{\pc}, assuming that a fraction $\varepsilon = 10^{-3}$ of the BH mass is accumulated in the axion cloud. Tab.~\ref{tab:ExpParas} summarizes the experimental parameters used for MAGGIE.
\begin{table}[h]
   \centering
   \begin{tabular}{lcc}
       \toprule
       \textbf{Symbol} & \textbf{Description} & \textbf{Value}\\
       \midrule
       $f$ & Frequency Range & \SI{10}{\kHz} - \SI{40}{\MHz} \\
       $B_0$ & Magnetic Field Strength & \SI{14}{\tesla}  \\
       SNR & Signal-to-Noise Ratio & 3 \\
       $N$ & Number of the loops & 10 \\
       $A_l$ &  Area of a single loop &  $\SI{914}{\cm^2}$ \\
       $T_{sys}$ & System Noise Temperature & \SI{350}{\kelvin} \\
       \bottomrule
   \end{tabular}
   \caption{Overview of the experimental parameters used for the proposed sensitivity in Figs.~\ref{fig:sens-continuous} and \ref{fig:sens-transient}.}
   \label{tab:ExpParas}
\end{table}

\subsection{Transient Signal Sensitivity}
\label{subsec:transient_sensitiity}

Beyond continuous searches, MAGGIE is also sensitive to transient HFGWs such as those from PBH mergers. Rather than treating the merger as a single narrowband signal at $f_{\mathrm{isco}}$, we construct a full inspiral-merger-ringdown template bank and perform a matched-filter analysis: A single reference waveform is generated with a standard Inspiral-Merger-Ringdown (IMR) model using \texttt{pycbc} \cite{pycbc} and mapped across the experiment's frequency band via the exact general relativity scale-freedom of the two-body problem used by IMRPhenomD \cite{Husa:2015iqa,Khan:2015jqa}. Rescaling the total mass by $\lambda$ at fixed mass ratio rescales time and hence frequency by $1/\lambda$ and amplitude by $\lambda$. So one calibration waveform generates the entire bank without repeated waveform generation.

\begin{figure}[t!]
    \centering    \includegraphics[width=0.7\linewidth]{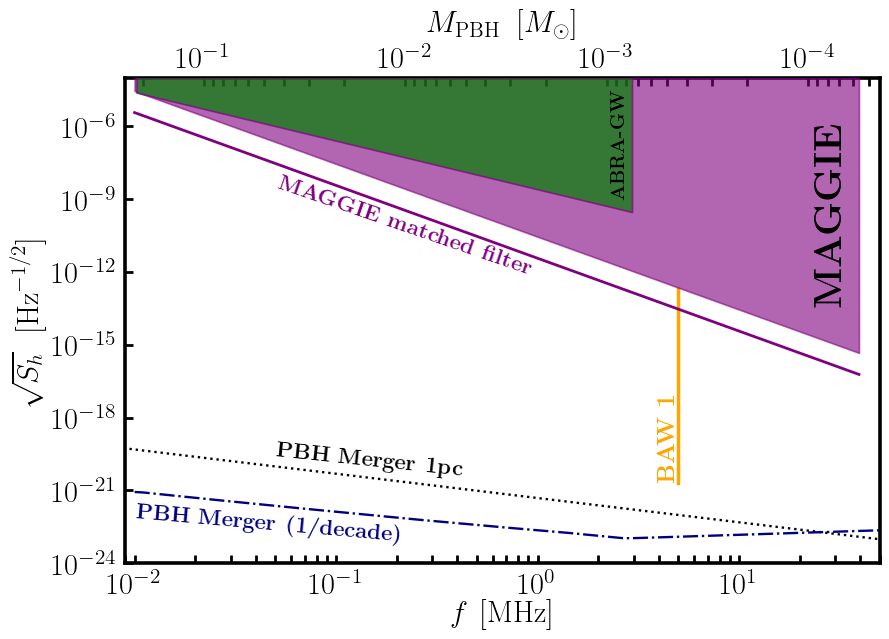}
    \caption{Projected strain-equivalent SND of MAGGIE for $\times$-polarized transient GW signals from equal-mass PBH binaries evaluated for $\text{SNR}=3$ (purple area) and the matched-filter sensitivity with a gain factor of $G\approx7.72$ (see Appendix~\ref{app:template} for details) compared to the narrowband estimate. The corresponding PBH mass is obtained from the ISCO relation \cite{Aggarwal_2021}. Using matched filtering, the expected sensitivity (purple line) is increased by approximately one order of magnitude. The latest ABRACADABRA limit \cite{Pappas_2025_Abra} and the BAW 1 limit \cite{Goryachev:2021zzn} are shown for comparison along with the expected strain from PBH mergers at a reference distance of \SI{1}{\pc} \cite{maggiore_2007}. Also shown are the strain sensitivities required to observe an individual PBH-binary merger at an expected rate of one per decade, following \cite{GWDomcke}. Experimental parameters used for this projection are summarized in Table~\ref{tab:ExpParas}.}
    \label{fig:sens-transient}
\end{figure}

Each filter is passed through the induced voltage transfer function of Eq.~\ref{eq:Uind} and compared against the system noise temperature $T_{\mathrm{sys}}$ using the transient matched-filter SNR $\rho$ of Eq.~\ref{eq:SNRMF}, integrated over each template's full bandwidth rather than a single frequency bin. By varying the distance of the source for each filter until a threshold of $\rho=3$ is exceeded, we obtain the corresponding strain SND shown as the purple line in Fig.~\ref{fig:sens-transient}. For comparison, the purple area in Fig.~\ref{fig:sens-transient} shows the corresponding strain-equivalent SND from Eq. \ref{eq:strain_sensitivityCS}. The two curves differ by a single, frequency-independent gain factor $G\approx7.72$ (see Appendix~\ref{app:template} for details) which corresponds to the sensitivity improvement matched filtering provides over an equivalent single-bin search. Tab.~\ref{tab:ExpParas} summarizes the experimental parameters used. A more detailed description of the matched-filter analysis is given in Appendix~\ref{app:TSD} and will also be the subject of an upcoming paper.

Computational constraints impose additional practical limits on sensitivity \cite{Cezar_2026_modular}. As detailed in Sect.~\ref{sec:daq}, the matched-filtering approach requires intensive processing to cover a broad signal parameter space. While our GPU-accelerated pipeline achieves promising performance \cite{Cezar_2026_modular}, ongoing optimization is essential to extend the bandwidth and duration of real-time data analysis without sacrificing sensitivity.

Finally, the same template catalog used above, generated from a single reference waveform and rescaled across the experiment's frequency band via the general relativity scale-freedom construction, is used in the signal-injection loop to verify the fidelity and timing of signal transfer through the full detector chain, ensuring accurate end-to-end system characterization.

\section{Discussion}
\label{sec:discussion}

In the broader HFGW experimental landscape, the Fermilab Holometer \cite{Holometer} utilizes correlated interferometry to probe the \si{\MHz} regime, while resonant microwave cavities like MAGO \cite{MAGO} plan to achieve high sensitivity within narrow frequency bands by searching for GW-induced mode excitations. Magnetic Weber bars detect periodic modulations of a background magnetic field through mechanical oscillations \cite{Domcke_2025_weberbar}, though they face challenges with vibrational isolation and coherence. Furthermore, collaborative initiatives such as GRAVNET \cite{GRAVNET} are emerging to correlate multiple high-frequency detectors globally, improving overall sensitivity and noise rejection. Finally, the US-based ABRACADABRA experiment \cite{AbraMain}, originally designed for axion searches, incorporates a broadband GW detection channel based on concepts closely related to MAGGIE \cite{Pappas_2025_Abra}, operating in a similar \SI{10}{\kHz} to \si{\MHz} range. Complementing these efforts, MAGGIE establishes the first dedicated lumped-element HFGW detector in Europe, providing an independent platform to cross-correlate high-frequency signals globally. Fig.~\ref{fig:sens-transient} compares the current and sensitivities of these experiments alongside characteristic strain predictions from sources such as PBH mergers.

Leveraging infrastructure from axion haloscopes, such as a \SI{14}{\tesla} solenoidal magnet, MAGGIE benefits from a favorable scaling of the induced voltage signal with frequency, $V_{\text{ind}} \sim \omega^3$. Meanwhile, the system noise remains approximately white across a wide band, resulting in improved strain sensitivity at \si{\MHz} frequencies. A first prototype of the MAGGIE detector, designed to probe the \SI{10}{\kHz} to \SI{40}{\MHz} band, is currently under construction at the University of Hamburg. Before deployment in the \SI{14}{\tesla} magnet, this prototype will undergo baseline characterizations of RF impedance, stray capacitance, and DAQ noise floor in a shielded test environment to validate the transfer functions. While the accessible frequency band is theoretically unbounded down to DC, the ultimate high-frequency limit is governed by the parasitic capacitance of the pickup loops and readout electronics.

\section{Outlook and Conclusion}
\label{sec:outlook_conclusion}

To further improve sensitivity, several hardware upgrades are planned for future iterations. The transition of the warm-bore setup to a fully cryogenic environment will lower the thermal noise floor. Simply cooling the system down to $\sim\SI{4}{\kelvin}$ reduces the system noise temperature $T_{\text{sys}}$ by about a factor of 100, which immediately yields an order-of-magnitude improvement in the strain sensitivity. Coupling this cryogenic environment with a superconducting readout system, such as Superconducting Quantum Interference Device (SQUID) amplifiers, could push the electronic noise floor even closer to the quantum limit. However, implementing current-sensitive SQUIDs alters the favorable $\omega^3$ voltage scaling from the high-impedance voltage detection scheme, requiring a trade-off between achieving quantum-limited noise and maintaining optimal high-frequency signal coupling.

In the lower-frequency regime (below \SI{10}{\MHz}), the sensitivity can be significantly enhanced by introducing a high-permeability ferrite core into the pickup loop. Similar to techniques used in lumped-element axion searches, a ferromagnetic material concentrates the magnetic flux generated by the GW-induced effective current. Since the measured flux is multiplied by the material's relative permeability $\mu_r$, this modification could boost the signal power and improve the strain sensitivity by several orders of magnitude.

On the data acquisition side, GPU-accelerated matched filtering is required to handle the large data throughput and enable real-time transient detection across dense parameter spaces. This data pipeline, along with the detailed performance of the transient search, will be covered in an upcoming paper. Furthermore, the broadband loop can be complemented by a tunable high-Q resonant LC circuit optimized for continuous-source operation, enabling targeted sensitivity enhancements and narrowband scanning within selected frequency ranges.

In conclusion, MAGGIE establishes a novel, scalable approach to HFGW detection by exploiting electromagnetic signals induced by GWs in a strong solenoidal magnetic field. With broadband sensitivity down to strain amplitudes of about $10^{-19}$ at \SI{40}{\MHz}, this concept opens a new experimental window into the \si{\MHz} regime. Upper limits on HFGWs can also be reinterpreted as constraints on the abundance of PBHs, translating non-detections into bounds on their contribution to dark matter (DM). A detailed derivation of this exclusion limit, utilizing local DM density constraints to bound the maximum allowed strain from nearby sources, is provided in Appendix~\ref{app:ExclDM}. Unlike microlensing surveys such as OGLE \cite{OGLE} or MACHO \cite{MACHO}, which constrain PBHs assuming Galactic-scale distances and smooth halo distributions, GW searches remain sensitive to nearby populations where local overdensities or clustering may evade lensing bounds. Focusing on distances of order a few parsecs therefore enables MAGGIE-like experiments to probe an otherwise unconstrained region of the PBH DM parameter space.

\ack{This project is funded by the Deutsche Forschungsgemeinschaft (DFG, German Research Foundation) under Germany’s Excellence Strategy – EXC 2121 ``Quantum Universe" – 390833306, and through the DFG funds for major instrumentation grant DFG INST 152/824-1. This article is based upon work from COST Action COSMIC WISPers CA21106, supported by COST (European Cooperation in Science and Technology).}

\bibliography{refs}
\clearpage

\appendix

\section{Transient sensitivity derivation}
\label{app:TSD}

\subsection{Strain-equivalent SND $\sqrt{S_h}$ of the detector}

To characterize the detector sensitivity in terms of the GW strain, we relate the experimentally measured voltage SND $\sqrt{S_V}$ to the strain-equivalent SND $\sqrt{S_h}$. This requires combining the detector response to a GW with the measured voltage-noise floor. To derive the transient sensitivity, the transfer function given in Eq.~\ref{eq:phifinal}, which describes the magnetic flux induced by a passing GW, must therefore be generalized to a transient chirp signal. The harmonic contribution of a chirp-like signal at a reference frequency $f_0$ to the induced magnetic flux can be written as:
\begin{equation}
    \phi_{\mathrm{ind}}(t) = \alpha \, \omega_0^2 \, h_{0} \, \sin(\omega_0 t),
    \label{eq:phigen}
\end{equation}
where $\alpha=\SI{2.21e-19}{\meter^2\second^2\tesla}$ characterizes the detector geometry and incorporates the external magnetic field $B_0$, the loop cross-sectional area $A_l$, and the number of windings $N$. 

The induced voltage $V_{\mathrm{ind}}(t)=-\partial_t\phi_{\mathrm{ind}}(t)$ is related to the frequency response through its Fourier transform:
\begin{equation}
    \begin{aligned}
        \tilde V(\omega) &=\mathcal{F}\{V_\mathrm{ind}(t)\} = \mathcal{F}\{-\partial_t\phi_{\mathrm{ind}}(t)\}
        \\ &=-i \, \omega \mathcal{F}^*\{\phi_{\mathrm{ind}}(t)\}
        \\ &= i  \, \alpha \, \omega_0^2 \, h_{0} \, \omega \, \delta(\omega-\omega_0).
    \end{aligned}
\end{equation}
The imaginary unit represents the physical $90^\circ$ phase shift between the incoming GW and the measured voltage. Averaging both sides over the spectral resolution bandwidth $f_s$ and the number of averages $N$ then relates the corresponding SNDs:
\begin{equation}
    \begin{aligned}
    \Rightarrow \frac{\tilde V(\omega_0)}{\sqrt{f_s N}} &= \, \alpha \, \omega_0^3 \, \frac{h_{0}}{\sqrt{f_s N}}
    \\ \Rightarrow \sqrt{S_V(\omega_0)}&= \, \alpha \, \omega_0^3 \sqrt{S_h(\omega_0)}.
    \end{aligned}
\end{equation}
Generalizing this relation to all frequency components $\omega$ yields the strain-equivalent SND of the detector:
\vspace{0.25cm}
\begin{equation}
    \sqrt{S_h(\omega)}= \frac{\sqrt{S_V(\omega)}}{\alpha \, \omega^3}
    \label{eq:SConversion}
\end{equation}

\subsection{Matched filtering}
\label{app:matchedfilter}
Having characterized the detector noise, we can now evaluate the expected signal strain from the induced magnetic flux. In the frequency domain, this follows directly from the Fourier transform of the time derivative of Eq.~\ref{eq:phigen}:
\begin{equation}
    \tilde V(\omega) =\mathcal{F}\{V_{\mathrm{ind}}(t)\} = \alpha \, \omega^3 \, \tilde h(\omega) \hspace{0.4cm} \Rightarrow \hspace{0.4cm} \tilde{h}(\omega) = \frac{ \tilde V(\omega)}{\alpha \, \omega^3}
\end{equation}

Using this transformation, we can evaluate the SNR relation from Eq.~\ref{eq:SNRMF_final}. Substituting the expressions for $\sqrt{S_h}$ and $\tilde h$ yields:
\begin{equation}
\begin{aligned}
    \rho_{\mathrm{mismatch}}(t) &= \underbrace{\frac{4}{\sqrt{\langle h_{\mathrm{t}}| h_{\mathrm{t}}\rangle}}}_{=:A} \, \Re\left\{ \mathcal{F}^{-1}\left\{\frac{\tilde h_\mathrm{t}}{S_h} \right\}\star h_{\mathrm{meas}} \right\}(t) \\
    &= A  \int \mathrm{d}\omega \, \frac{\tilde h_t^*(\omega) \, \tilde h_{\mathrm{meas}}(\omega)}{S_h(\omega)} e^{i \omega t} =A \int \mathrm{d}\omega \frac{\tilde h_t^*(\omega) \, \tilde h_{\mathrm{meas}}(\omega) \, \alpha^2 \, \omega^6}{S_V(\omega)} e^{i \omega t} \\
    &=A \int \mathrm{d}\omega \frac{\tilde V_t^*(\omega) \, \tilde h_{\mathrm{meas}}(\omega) \alpha \, \omega^3}{S_V(\omega)} e^{i \omega t}=A \int \mathrm{d}\omega \, \frac{\tilde V_t^*(\omega) \, \tilde V_\mathrm{meas}(\omega)}{S_V(\omega)}e^{i \omega t} \\
    &= \frac{4}{\sqrt{\langle h_{\mathrm{t}}| h_{\mathrm{t}}\rangle}} \, \Re\left\{\mathcal{F}^{-1}\left\{\frac{\tilde V_t}{S_V} \right\}\star V_\mathrm{meas} \right\}(t) \geq 3
\end{aligned}
\end{equation}
This result is the cross-correlation of the detector's voltage output in the time domain, $ V_\mathrm{meas} (t)$, with the inverse Fourier Transform of the optimal filter in units of volts $\mathcal{F}^{-1}\left\{\frac{\tilde V_t}{S_V} \right\} = \mathcal{F}^{-1}\left\{\frac{\tilde h_t(\omega)\, \alpha \, \omega^3}{S_V} \right\}$. The voltage noise power spectral density $S_V$ needs to be estimated experimentally and, in the following, is taken to be the thermal noise limit of the experiment. The cross-correlation is performed over the sampling range of MAGGIE, i.e., $f\in [0.01,40] \, \SI{}{\mega \hertz}$, and we require a minimum SNR of $\rho \geq 3$.

This result is the cross-correlation of the detector's voltage output in the frequency domain, $\tilde V_\mathrm{meas}(\omega)=\tilde h_\mathrm{meas}(\omega)\, \alpha \, \omega^3$, with the voltage signal template $\tilde V_t = \tilde h_t(\omega)\, \alpha \, \omega^3$. 

\subsection{Template bank and evaluation}
\label{app:template}

All chirp templates for the matched-filtering analysis are created with the \texttt{pycbc} package \cite{pycbc}. For the presented exclusion limit, the parameter space was reduced to two free parameters $[M_\mathrm{tot}, d]$, the total mass $M_\mathrm{tot}$ and the distance $d$. All other parameters, such as spin, eccentricity, and inclination, are fixed. For an equal mass ratio, one template in a calibrated mass range of $M_\mathrm{tot}=20 \mathrm{M}_\odot$ was created and then scaled down using the scale freedom of general relativity \cite{Husa:2015iqa,Khan:2015jqa}.

We start with a first voltage signal template $V_{t, \, j=1}$ and cross-correlate with a voltage output of the detector $V_{\mathrm{meas}, \, i=1}$ while varying the distance parameter of $V_{\mathrm{meas}, \, i=1}$ until the SNR $\rho$ exceeds $3$. Then the process is repeated with the next detector output $V_{\mathrm{meas}, \, i=2}$ until all templates covering the targeted frequency range have been cross-correlated. Then the next voltage signal template $V_{t, \, j+1}$ is taken, and the whole process is repeated.

The resulting minimal detectable strain for the SNR threshold of $\rho=3$ is converted to its strain-equivalent SND, $\sqrt{S_h^{\mathrm{matched \,filter}}}$, and compared against the detector's strain-equivalent SND for a narrowband SNR evaluation (Eqs.~\ref{eq:SConversion} and \ref{eq:strain_sensitivityCS}), leading to a gain factor
\begin{equation}
	G = \sqrt{\frac{S_h}{S_h^{\mathrm{matched\, filter}}}} \approx 7.72 .
\end{equation}
The projected strain-equivalent SND for the matched-filtering approach, including the gain $G\approx7.72$ from our sensitivity analysis, is shown as the purple line in Fig.~\ref{fig:sens-transient}. This plot reflects the envelope of the algorithm introduced above and is dominated by the cross-correlation where the signal and detector templates are equal.

\section{Exclusion from Dark Matter Density constraints}
\label{app:ExclDM}

An exclusion limit on the GW strain can be derived from superradiant sources based on constraints from the local DM density. We begin by assuming that PBHs constitute a fraction of the local DM density. Observations of galactic dynamics constrain the DM density in the solar neighborhood to be approximately
\begin{equation}
    \rho_{\rm DM,local} \lesssim 0.01\, M_\odot\,\mathrm{pc}^{-3},
\end{equation}
up to $\mathcal{O}(1)$ uncertainties depending on the assumed halo model and baryonic contributions. See \cite{DMdensity,2026arXiv260504801H} for local DM density estimates.

For a single mass source $M_{\rm BH}$ located at a distance $d$, we can estimate its effective contribution to the local DM density by distributing its mass over a volume $\propto d^3$. This yields:
\begin{equation}
    \rho \propto \frac{M_{\rm BH}}{d^3} \lesssim \rho_{\rm DM,local}.
    \label{eq:densityconstrain}
\end{equation}
This relation can be interpreted as a consistency condition. The presence of a single nearby object should not exceed the local DM density.

Consequently, this inequality sets a lower bound on the distance to an object with mass $M_{\mathrm{BH}}$. Physically, it implies that sufficiently massive objects cannot be arbitrarily close together, as this would locally violate the observed DM density.

We can now connect the density constraint derived above to the expected signal from black hole axion superradiance. The characteristic axion mass is set by the condition that its Compton wavelength is comparable to the size of the horizon of the black hole. This yields the scaling relation
\begin{equation}
    m_a \sim \left(\frac{M_\odot}{M_{\rm BH}}\right) 10^{-10}\,\mathrm{eV},
\end{equation}
which follows from the requirement that the gravitational radius matches the axion wavelength \cite{Arvanitaki1}.

The emitted GWs arise dominantly from axion annihilation processes and have a frequency set by twice the axion mass. This leads to:
\begin{equation}
    f_{\rm GW} \sim 2 \left(\frac{m_a}{10^{-9}\,\mathrm{eV}}\right) 10^6\,\mathrm{Hz},
\end{equation}
where the numerical scaling reflects the conversion between energy and frequency \cite{Aggarwal_2021}. Combining these relations allows one to eliminate the axion mass and express the black hole mass directly in terms of the observable GW frequency:
\begin{equation}
    M_{\rm BH}(f_{\rm GW}) \approx 0.2\,M_\odot \left(\frac{10^6\,\mathrm{Hz}}{f_{\rm GW}}\right).
\end{equation}
This relation provides a direct mapping between the source mass scale and the frequency band probed by the detector, and will be used in the following to translate density constraints into bounds on the observable strain.

Inserting the mass-frequency relation into Eq.~\ref{eq:densityconstrain}, we can express the minimal allowed distance directly in terms of the GW frequency:
\begin{equation}
 d\gtrsim \left( \frac{M_{BH}}{0.01 M_\odot}\right)^{1/3} \,\mathrm{pc} = \left(\frac{0.2}{0.01}\right)^{1/3}\left(\frac{10^6\,\mathrm{Hz}}{f_{\rm GW}}\right)^{1/3} \,\mathrm{pc} \approx 2.7 \, \left(\frac{10^6\,\mathrm{Hz}}{f_{\rm GW}}\right)^{1/3} \,\mathrm{pc}.
\end{equation}
The characteristic strain from superradiant emission can be estimated from the GW power emitted by the axion cloud surrounding the black hole:
\begin{equation}
    h \sim 10^{-18} \left(\frac{1\,\mathrm{pc}}{d}\right)\left(\frac{M_{\rm BH}}{2 M_\odot}\right).
\end{equation}
This scaling reflects that the emitted amplitude is set by a fraction of the black hole mass stored in the boson cloud \cite{Arvanitaki2}. Substituting the frequency-dependent expressions for $d$ and $M_{\rm BH}$ derived above, we obtain an exclusion curve of the form:
\begin{equation}
    h< 10^{-18} \frac{1 \, \mathrm{pc}}{2.7 \, \mathrm{pc}} \left( \frac{f_{\rm{GW}}}{10^6\,\mathrm{Hz}} \right)^{1/3} \frac{0.2}{2} \left(\frac{10^6\,\mathrm{Hz}}{f_{\rm{GW}}}\right) = 3.7\times10^{-20} \left(\frac{f_{\rm GW}}{10^6\,\mathrm{Hz}}\right)^{-2/3}
\end{equation}
\begin{equation}
    h_{\rm excl}(f) \propto f^{-2/3}.
\end{equation}
This relation defines the maximum allowed strain as a function of frequency, assuming that PBHs saturate the local DM density.

\end{document}